# Deformations of one-dimensional block media

N.I. Aleksandrova

*N.A. Chinakal Institute of Mining, Siberian Branch, Russian Academy of Sciences,*

*Krasnyi pr. 54, Novosibirsk, 630091, Russia,*

*e-mail: alex@math.nsc.ru*

**Abstract** The paper gives a description of wave propagation in discrete-periodic one-dimensional media with block structure. For one-dimensional problems mathematical models are proposed that describe block structures in the form of a mass chain or bars connected by elastic springs and viscous dampers. For these models, the numerical calculations of the parameters of perturbations are obtained as well as asymptotic solutions at large time since the beginning of pulse action. The numerical calculations and analytical solutions are compared.

**1. Introduction**

According to the common geophysical and geomechanical concept, the rock mass is a uniform medium and its deformation is described by the well-developed linear theory of elastic wave propagation. On the basis of this theory are constructed the methodological basis for calculating the stress state of rocks near the mines, the seismic processing as well as the interpretation of seismic data in geophysics and mining.

A serious reason to revise the prevailing concept is provided by investigations made during recent years. They indicate the need to consider the block structure of rocks in the construction of mathematical models of geomechanical and seismic phenomena. The development of these research was greatly influenced by the Sadovskiy's fundamental concept of the block-hierarchical structure of geological objects [1]. According to this concept a rock mass is treated as a system of nested blocks of different scale levels connected by intermediate layers consisting of weaker fractured rocks [2]. Both in statics and dynamics, the presence of such intermediate supple layers leads to the fact that the deformation of the block rock masses is mainly due to the deformation of the layers.

Experimental and theoretical studies of wave propagation in one-dimensional models of block media are given in [2–15]. These articles show that the representation of blocks as massive non-deformable bodies makes it possible to select from a complex dynamic behavior of the elastic block medium that part which is determined by the deformation of the layers between the blocks. In the wave of deformation caused by a shock load, low-frequency waves of the pendulum-type appear [2].

Distinctive features of the pendulum waves are their relatively long duration though the effect was short-pulse, low velocity of propagation that is much lower than the velocity of the elastic wave in the rock blocks, and low attenuation [2–7].

Based on the experimental data from [2–7], in this paper we treat the block medium as a one-dimensional discrete-periodic structure. Thus appearing models are simple enough and allow us to obtain analytical solutions, which are presented below.

## 2. Transient waves in the chain of masses and springs

General regularities inherent in the waves in block rock masses can be found on the models of a discrete periodic structure. Let us first analyze the simplest model of a block structure. We replace each block by a rigid ball with mass $m$ equal to the mass of the block and replace each intermediate layer by a spring. As a result, we obtain a coupled system of pairwise equal spring pendulums. Waves propagating in such system are called pendulum waves [2].

Motion equations of the mass-and-spring chain have the following form:

$$m\ddot{u}_n = k(u_{n+1} - 2u_n + u_{n-1}) + P_0 \delta_{0n} Q(t); \quad n = -\infty, ..., \infty \tag{1}$$

where $u_n$ is the displacement of the $n$-th mass and $k$ is the stiffness of a spring, $\delta_{0n}$ is the Kronecker delta function. Let, at time $t = 0$, the zero mass experiences a half-sine pulse force $Q(t)$ with the amplitude $P_0$ as a simulation of an impulsive load:

$$Q(t) = \sin(\omega_* t) H(\pi - \omega_* t) H(t), \tag{2}$$

where $\omega_*$ is the impulsive frequency, and $H$ is the Heaviside function. Initial conditions are zero: $u_n = 0$, $\dot{u}_n = 0$ at $t = 0$.

Apply the Laplace transform with respect to time to (1) and (2) with zero initial conditions. Then apply the discrete Fourier transform with respect to the axial coordinate $n$.

The transformed solution has the form:

$$u^{LF_d} = \frac{Q^{LF_d}}{mD(p,q)}, \quad Q^{LF_d}(p,q) = \frac{P_0(1+e^{-p\pi/\omega_*})\omega_*}{\omega_*^2 + p^2}, \quad D(p,q) = p^2 + 4\frac{k}{m}\sin^2\frac{ql}{2}, \tag{3}$$

where $D(p,q)$ is the dispersion operator of the system, $L$ symbolizes the Laplace transform, $p$ is the Laplace transform parameter, $F_d$ stands for the discrete Fourier transform, $q$ is the discrete Fourier transform parameter, and $l$ is the length of a spring.

It is impossible to express the inverse Laplace transform of (3) in terms of the known functions. Hence, we try to find an asymptotic solution describing a long-wave disturbance ($|q| \ll 1$) for large time since the beginning of the process ($t \to \infty$; note that the condition $t \to \infty$ corresponds to the condition $p \to 0$). For this we use the method, suggested by Slepyan [9], for joint inversion of the

Laplace–Fourier images in a neighborhood of the half line $nl = c_* t$, where $c_* = l\sqrt{k/m}$ is the velocity of the propagation of infinitely long waves. As a result we obtain the following asymptotic solution:

$$\Delta u_n = \frac{u_{n+1} - u_{n-1}}{2} \approx -\frac{l}{c_*}\frac{\partial u_n}{\partial t} \approx -\frac{2P_0 l^2}{mc_*\omega_*(\gamma t)^{1/3}}\mathrm{Ai}(\kappa), \quad \frac{\partial^2 u_n}{\partial t^2} \approx -\frac{2P_0 l c_*}{m\omega_*(\gamma t)^{2/3}}\frac{d\mathrm{Ai}(\kappa)}{d\kappa}, \tag{4}$$

$$\gamma = \frac{c_* l^2}{8}, \quad \kappa = \frac{nl - c_* t}{(\gamma t)^{1/3}}, \quad \mathrm{Ai}(\kappa) = \frac{1}{\pi}\int_0^\infty \cos(\kappa z + z^3/3)\,dz,$$

where $\mathrm{Ai}(\kappa)$ is the Airy function [16].

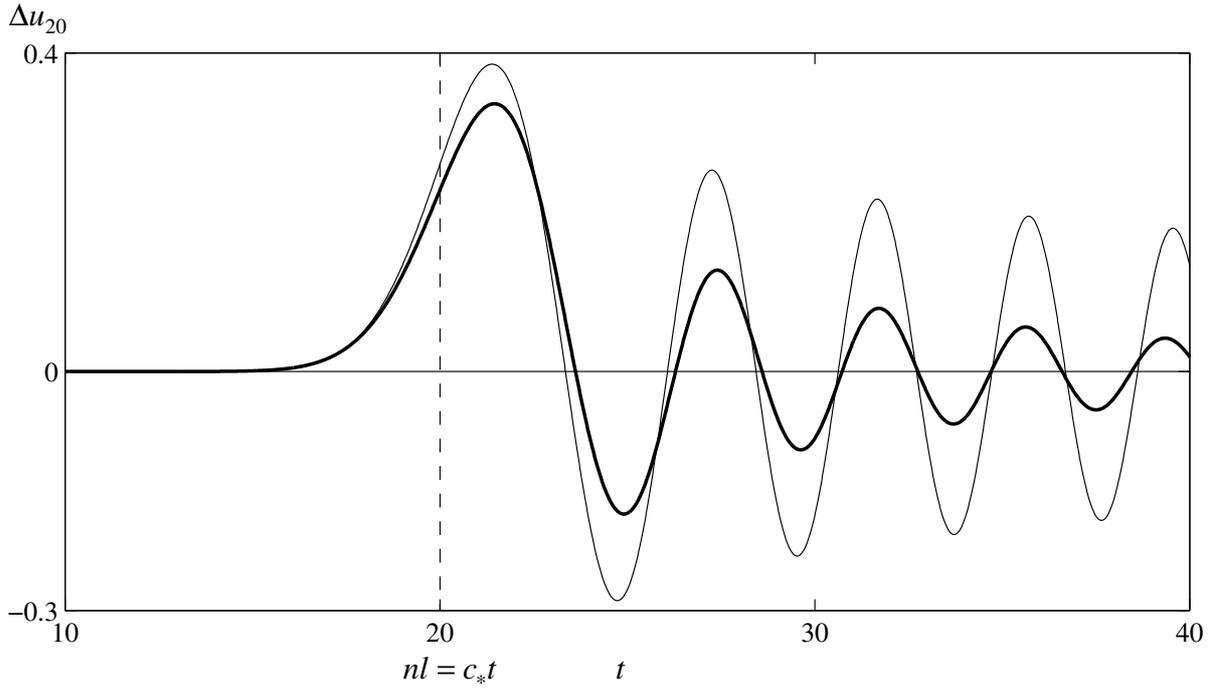

**Fig. 1.** The asymptotic and numerical solutions under impulsive input ($n = 20$).

From (4) it follows that, with increasing time, the strain amplitudes of the springs and the velocity amplitudes of the masses decrease in proportion to $t^{-1/3}$, the acceleration amplitude drops in proportion to $t^{-2/3}$, and the quasi-front zone expands in proportion to $t^{1/3}$. The asymptotic behavior of the strains of the springs as well as of the velocities and accelerations of the masses is determined by the Airy function.

The time versus strain curves in Fig. 1 are obtained for $n = 20$, $m = 1$, $l = 1$, $k = 1$, $\omega_* = 2$ analytically from (4) (thin curve) and numerically by a finite-difference method (thick curve). The vertical dotted line indicates the position of the quasi-front: $nl = c_* t$. Near the quasi-front, the asymptotic solution (4) agrees with the finite difference solution of (1) both qualitatively and quantitatively.

## 3. Bar model

Let us analyze the propagation of waves in more complex models of block media. Consider a system of identical elastic bars connected by identical springs. The bars have the following parameters: $\rho$ is the density, $c$ is the sound speed, $l$ is the length, and $S$ is the area of the cross-section. The motion equations of the bars are as follows:

$$\ddot{u}_n = c^2 u_n'', \quad n = 0,...,\infty, \tag{5}$$

where $u_n$ is the displacement of the $n$-th bar, the stroke means the derivative with respect to the $x$-coordinate. The stresses on the ends of the bars are supposed to be proportional to the elongation of the springs:

$$SEu_n'\big|_{x=l(n-1/2)} = -k\left(u_{n-1}\big|_{x=l(n-1/2)} - u_n\big|_{x=l(n-1/2)}\right), \quad n = 1,...,\infty; \tag{6}$$

$$SEu_n'\big|_{x=l(n+1/2)} = -k\left(u_n\big|_{x=l(n+1/2)} - u_{n+1}\big|_{x=l(n+1/2)}\right), \quad n = 0,...,\infty;$$

$$Eu_0'\big|_{x=-l/2} = P_0 Q(t).$$

Here $E$ stands for the Young modulus. Initial conditions are zero. The half-sine pulse (2) with the amplitude $P_0$ is applied to the left end of the zero bar.

In order to obtain an analytical solution, we suppose that the discrete system is infinite ($-\infty \leq n \leq \infty$) and the impulsive load (2) is applied at the center of the zero bar (for symmetry). Apply the Laplace transform with respect to time to (5) and (6) endowed with the zero initial conditions. Then apply to (5) the continuous Fourier transform with respect to the longitudinal coordinate $x$.

The transformed solution for the bar system is similar to (3):

$$u^{LF} = \frac{P_0 Q^L C(p,q)}{E D(p,q)}, \quad D(p,q) = \cos ql - \cosh\frac{pl}{c} - \frac{p}{2\alpha}\sinh\frac{pl}{c}, \quad \alpha = \frac{kc}{FE}. \tag{7}$$

Now compare the dispersion properties of the two models described. In the chain of masses model, the phase and group velocities are as follows:

$$c_f = \frac{\omega}{q} = \frac{2}{q}\sqrt{\frac{k}{m}}\sin\frac{ql}{2}, \quad c_g = \frac{d\omega}{dq} = l\sqrt{\frac{k}{m}}\cos\frac{ql}{2}. \tag{8}$$

Obviously, they are equal to each other when $q \to 0$:

$$c_f\big|_{q\to 0} = c_g\big|_{q\to 0} = l\sqrt{k/m} = c_*,$$

i.e. the long waves travel without dispersion and generate the quasi-front.

In the chain of bars model, the higher oscillation modes appear. They are associated with reflections from the ends of the bars. From (7) it is impossible to find an explicit dependence of $c_f$ and $c_g$ on the wave number $q$. Nevertheless, assuming the parameter $\beta = \alpha l/c$ small enough, one can derive explicit approximate formulas for all oscillation modes. Due to the periodicity of the trigonometric functions, there are infinitely many such modes. For the first and the second modes we have:

$$c_f^{\mathrm{I}} \approx \frac{2c_1}{ql}\sin\frac{ql}{2}, \quad c_g^{\mathrm{I}} \approx c_1\cos\frac{ql}{2}, \quad \omega^{\mathrm{I}} \approx \omega_0\sin\frac{ql}{2}, \tag{9}$$

$$c_f^{\mathrm{II}} \approx \frac{1}{q}\left(\omega_2 - \frac{4k}{\omega_2 m}\sin^2\frac{ql}{2}\right), \quad c_g^{\mathrm{II}} \approx c_2\sin ql, \quad \omega^{\mathrm{II}} \approx qc_f^{\mathrm{II}}, \tag{10}$$

$$\omega_0 \approx 2\sqrt{\frac{k}{m(1+\beta/3)}}, \quad \omega_1 = \frac{\pi c}{l}, \quad \omega_2 \approx \frac{\pi c}{l}\left(1 + \frac{4\beta}{\pi^2 + 4\beta}\right), \tag{11}$$

$$c_1 = l\sqrt{k/(m(1+\beta))}, \quad c_2 = 2\alpha c/\omega_2. \tag{12}$$

Qualitatively, the dispersion relations for the chain of masses (8) and for the first oscillation mode for the chain of bars (9) are similar to each other; quantitatively, they differ a little: the maximum values of $c_f$ and $c_g$ in the chain of bars are $\sqrt{1+\beta}$ times smaller than in the chain of masses and the maximum frequency in the chain of bars is $\sqrt{1+\beta/3}$ times smaller than in the chain of masses. The frequency $\omega_1$ (11) corresponds to oscillations with a period equal to the travel time of the longitudinal wave along the bar back and forth. It follows from (12) that $c_1 > c_2$ for all parameters of the problem, i.e., the high-frequency wave never outruns the low-frequency wave.

The asymptotics of the low-frequency longitudinal wave in the chain of bars is similar to (4), but has different coefficients:

$$\frac{\partial u}{\partial x} \approx \frac{P_0 c_*}{E\omega_*(\gamma t)^{1/3}}\mathrm{Ai}(\kappa), \quad \kappa = \frac{nl - c_* t}{(\gamma t)^{1/3}}, \quad \gamma = \frac{c_* l^2}{8(1+\beta)^2}, \quad c_* = c_1. \tag{13}$$

In order to compare the propagation processes of disturbances in the two models of the block media and find the limits of applicability of the asymptotic solutions, we carried out numerical calculations and compared the analytical estimates with the numerical results. The system of equations (5) with zero initial conditions and the boundary conditions (6) was solved using an explicit finite difference scheme. Propagation of perturbations in the chain of bars is illustrated in Fig. 2 for the case of the impulsive loading (2) with $\omega_* = 11.8\,\mathrm{ms}^{-1}$, where oscillograms of deformation $u'_n E/P_0$ are shown in the middle of the $n$-th bar ($n = 5$ and $n = 15$ in Fig. 2). The dashed lines show the quasi-fronts: $x = c_1 t$, $x = c_2 t$. The problem was solved numerically with the following parameters: $\rho = 7800\,\mathrm{kg/m^3}$; $l = 0.5\,\mathrm{m}$; $S = 0.00125\,\mathrm{m^2}$; $k = 48.75\,\mathrm{kg/m^2}$; $c = 5\,\mathrm{m/ms}$. These parameters correspond to: $m = 4.875\,\mathrm{kg}$ and $\beta = 0.1$. In Fig. 2, the thin lines illustrate the finite difference solutions for the chain of bars, while the thick lines illustrate the finite difference solutions

$\Delta u_n k /(P_0 S)$ for the chain of masses with the following parameters: $m = 4.875$ kg and $k = 48.75$ kg/m$^2$. The asymptotics of the long wave disturbances (13) for the chain of bars is not shown in Fig. 2 because, in the quasi-front zone, it is very close to the numerical results.

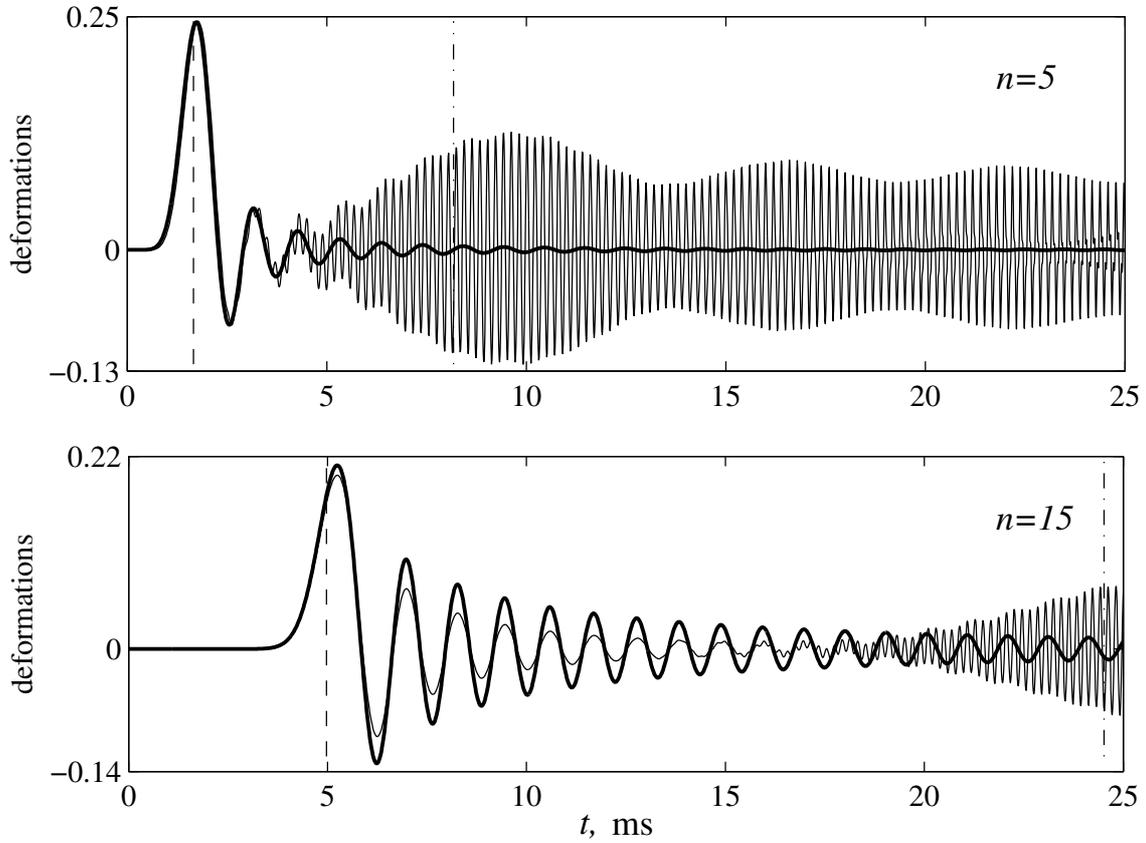

**Fig. 2.** The oscillograms of strains in the chains of masses and bars with $\beta = 0.1$.

Fig. 2 shows that the chain of asses model describes quite well the low-frequency perturbations in the chain of bars model. In the latter model, the high-frequency disturbance with frequency $\omega_1$ moves behind the long-wave disturbances, moving with velocity $c_1$, see (12). The velocity of the envelope of the high-frequency waves is close in value to the maximum group velocity of the second mode $c_2$, see (12).

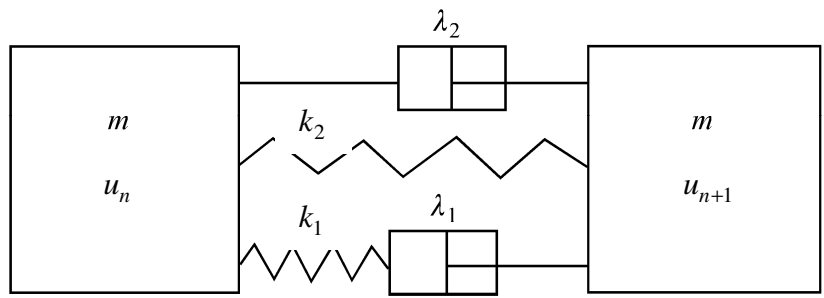

**Fig. 3.** The viscous-elastic model of intermediate layer deformation.

We made numerical calculations for various values of frequency $\omega_*$. The results of those calculations show that, as $\omega_*$ decreases, the low-frequency wave amplitude increases, while the high-frequency wave amplitude decreases.

So, we have demonstrated that the impulsive impact generates two groups of nonstationary waves in the chain of bars model. In order to obtain an approximation description ($\beta \ll 1$) of the low-frequency wave processes, we may replace a chain of bars by a chain of masses, such that the masses are equal to the masses of the bars. Using experimental data of frequencies and velocities of these waves in a block-structured medium, we can determine the structure and mechanical parameters of this medium.

### 4. Effect of viscosity of intermediate layers

To describe the behavior of layers between the bars we suggest the simultaneous use of two versions, parallel and serial, of the connection of viscous and elastic components.

Let us discuss a model, shown in Fig. 3, which is composed of elastic and damping elements joining-up stiff blocks of masses $m$. The layers between the masses are identical, the length of the springs is equal to $l$. The compression force $F$ is determined by the expression $F = k\delta$ for an elastic element and by the expression $F = \lambda \dot{\delta}$ for a damping element. Here $\delta$ and $\dot{\delta}$ stand for the element's elongation and the velocity of the element's elongation; $k$ and $\lambda$ stand for stiffness and coefficient of viscosity respectively.

If the elastic and damping elements are connected in series, the force of compression is given by the equation $\dot{F}/k_1 + F/\lambda_1 = \dot{\delta}$. Integrating yields

$$F = k_1 e^{-\alpha_1 t} \int_0^t e^{\alpha_1 t} \dot{\delta} dt, \quad \alpha_1 = k_1/\lambda_1.$$

If the elastic and damping elements are connected in parallel, the forces from all the elements are summed. As a result, in the model in Fig. 3, the force $F_n^r$ acting on the mass with the coordinate $n$ from the right-hand side is given by the expression

$$F_n^r = k_2(u_{n+1} - u_n) + \lambda_2(\dot{u}_{n+1} - \dot{u}_n) + k_1 e^{-\alpha_1 t} \int_0^t e^{\alpha_1 t} (\dot{u}_{n+1} - \dot{u}_n) dt.$$

There is a similar expression for the force $F_n^l$ acting from the left-hand side. Using the law of the motion $m\ddot{u}_n = F_n^r - F_n^l$, we obtain the following system of equations for the displacement of masses in the model shown in Fig. 3:

$$m\ddot{u}_n = k_2(u_{n-1} - 2u_n + u_{n+1}) + \lambda_2(\dot{u}_{n-1} - 2\dot{u}_n + \dot{u}_{n+1}) +$$
$$+ k_1 e^{-\alpha_1 t} \int_0^t e^{\alpha_1 t} (\dot{u}_{n-1} - 2\dot{u}_n + \dot{u}_{n+1}) dt, \quad n = 1, \ldots, N-1;$$

$$m\ddot{u}_0 = k_2(u_1 - u_0) + \lambda_2(\dot{u}_1 - \dot{u}_0) + k_1 e^{-\alpha_1 t} \int_0^t e^{\alpha_1 t}(\dot{u}_1 - \dot{u}_0)dt - P_0 Q(t), \tag{14}$$

$$m\ddot{u}_N = k_2(u_{N-1} - 2u_N) + \lambda_2(\dot{u}_{N-1} - 2\dot{u}_N) + k_1 e^{-\alpha_1 t} \int_0^t e^{\alpha_1 t}(\dot{u}_{N-1} - 2\dot{u}_N)dt.$$

In [6], it is shown that the spread of a low-frequency pendulum wave in the experimental block system under an impulse load can be numerically described using the one-dimensional model of the chain of masses with elastic springs and viscous dampers. The parameters $\lambda_1, \lambda_2, k_1, k_2$ of that viscoelastic model are chosen from the condition of matching the calculated and experimental values of the wave propagation velocity, its attenuation, and its characteristic period of oscillation.

An analytical solution describing the attenuation of pendulum waves can be obtained for the Voigt model of the viscoelastic deformation of layers between the blocks (Fig. 3, $k_1 = 0$). For such chain of masses, the motion equations are as follows:

$$m\ddot{u}_n = k_2(u_{n-1} - 2u_n + u_{n+1}) + \lambda_2(\dot{u}_{n-1} - 2\dot{u}_n + \dot{u}_{n+1}), \quad n = 1,...,N-1, \tag{15}$$

$$m\ddot{u}_0 = k_2(u_1 - u_0) + \lambda_2(\dot{u}_1 - \dot{u}_0) + P_0 Q(t), \quad u_N = 0.$$

For an infinite chain of masses, the Laplace–Fourier transformed solution is as follows:

$$u^{LF} = \frac{P_0 Q^L}{mD(p,q)}, \quad D(p,q) = p^2 + \frac{4}{m}\sin^2\left(\frac{ql}{2}\right)(\lambda_2 p + k_2).$$

The dispersion equation $D(i\omega, q) = 0$ implies that the phase and group velocities of the first mode on the dispersion curve are equal to each other when $q \to 0$ and coincide with the long-wave velocity $c_*$ when $\lambda_2 = 0$.

In the vicinity of the quasi-front $x = c_* t$, the asymptotic behavior of long-wave perturbations as $t \to \infty$ is given by the formulas

$$\ddot{u}_n = \frac{2P_0 c_* l}{m\omega_*(\gamma t)^{2/3}} \frac{1}{\pi} \int_0^\infty \sin(\kappa z + z^3/3) e^{-\mu z^2} z \, dz, \tag{16}$$

$$\dot{u}_n = \frac{2P_0 l}{\omega_* m(\gamma t)^{1/3}} \frac{1}{\pi} \int_0^\infty \cos(\kappa z + z^3/3) e^{-\mu z^2} dz,$$

$$\gamma = c_* l^2 \left(1 + 3\lambda_2^2/(mk_2)\right)/8, \quad \kappa = (nl - c_* t)/(\gamma t)^{1/3},$$

$$\mu = \alpha_1 t/(\gamma t)^{2/3}, \quad \alpha_1 = \lambda_2 l^2/(2m).$$

It follows from (16) that, when $\lambda_2 = 0$, there is no exponential attenuation ($\mu = 0$) and these formulas match with (4). The maximum amplitudes of the velocities and accelerations in the quasi-front zone monotonically decrease as $n \to \infty$:

$$\max_t \dot{u}_n \approx \frac{2.14 P_0}{m\omega_* n^{1/3}}, \quad \max_t \ddot{u}_n \approx \frac{2.07 P_0 c_*}{lm\omega_* n^{2/3}}. \tag{17}$$

The presence of viscosity leads to additional attenuation. When $\lambda_2 \neq 0$ and $n \to \infty$, (16) can be simplified:

$$\ddot{u}_n = \frac{4P_0 c_*}{\omega_* l \lambda_2 t} \frac{1}{\pi} \int_0^\infty \sin(\tilde{\kappa} z) e^{-z^2} z\, dz = \frac{P_0 c_*}{\omega_* l \lambda_2 \sqrt{\pi}} \frac{\tilde{\kappa}}{t} \exp\left(-\tilde{\kappa}^2/4\right),$$

$$\dot{u}_n = \frac{2P_0 l}{\omega_* m \sqrt{\alpha t}} \frac{1}{\pi} \int_0^\infty \cos(\tilde{\kappa} z) e^{-z^2} dz = \frac{P_0 l}{\omega_* m \sqrt{\pi \alpha t}} \exp\left(-\tilde{\kappa}^2/4\right),$$

(18)

$$\tilde{\kappa} = (nl - c_* t)/(\alpha t)^{1/2}.$$

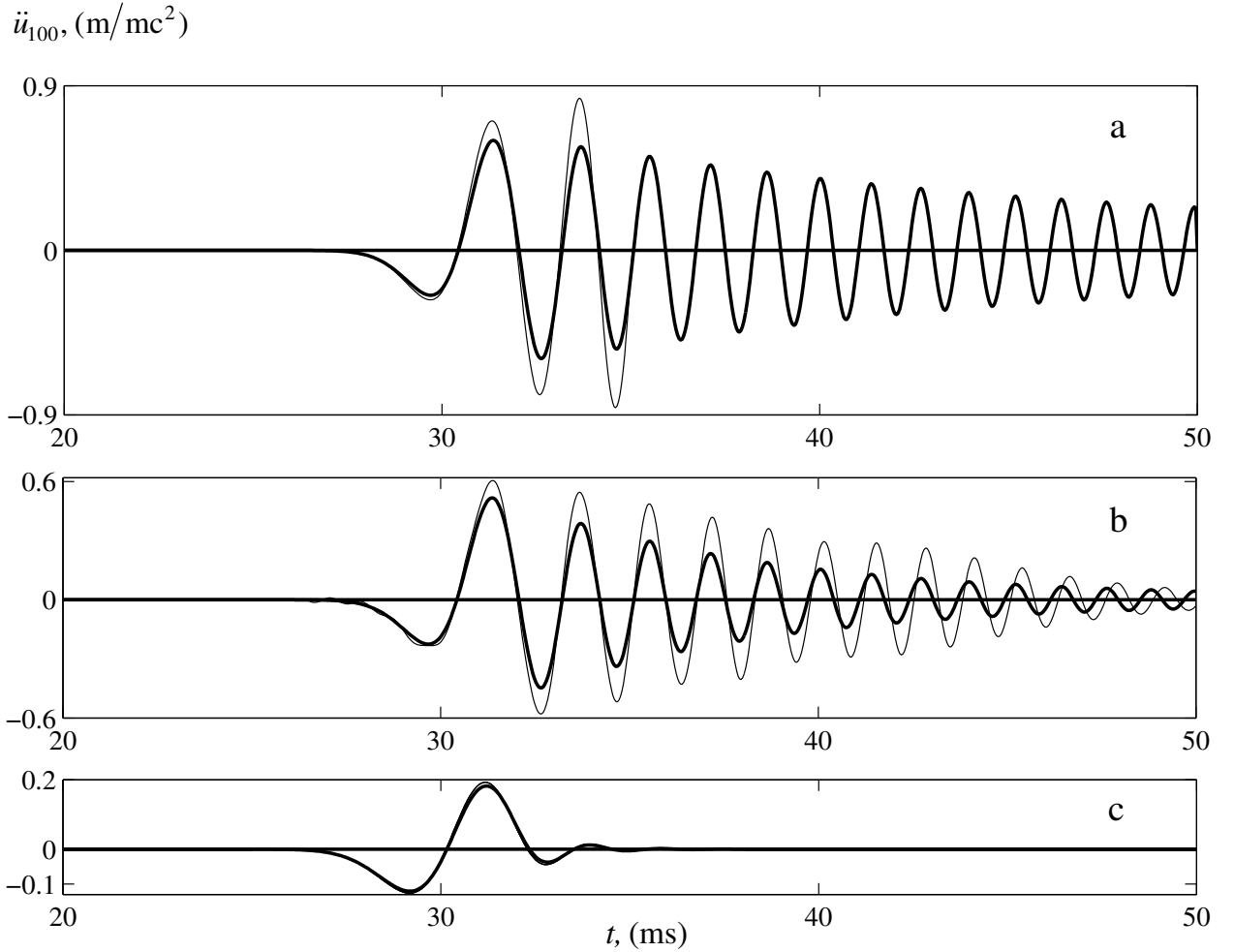

**Fig. 4.** The oscillograms of accelerations with different $\lambda_2$ ($n = 100$): (a) $\lambda_2 = 0$; (b) $\lambda_2 = 0.01$; (c) $\lambda_2 = 0.1$.

The maximum amplitudes of the velocities of masses decrease in proportion to $n^{-1/2}$ and the maximum amplitudes of the velocities of accelerations decrease in proportion to $n^{-1}$ ($n \to \infty$):

$$\max_t \dot{u}_n \to \frac{P_0}{\omega_*} \sqrt{\frac{2c_*}{\pi m l \lambda_2 n}}, \quad \max_t \ddot{u}_n \to \frac{P_0 k_2}{\omega_* m \lambda_2 n} \sqrt{\frac{2}{\pi e}}.$$

In Fig. 4, we plot the acceleration oscillograms for the $n$-th mass ($n = 100$) that are calculated with the use of the finite difference method (thick lines) and the asymptotic formulas (thin lines) for the following values of $\lambda_2$: $\lambda_2 = 0$ Ns/m, $\lambda_2 = 0.01$ Ns/m, and $\lambda_2 = 0.1$ Ns/m. More precisely, the asymptotic formula (16) is used for $\lambda_2 = 0$ Ns/m and $\lambda_2 = 0.01$ Ns/m, while the asymptotic formula (18) is used for $\lambda_2 = 0.1$ Ns/m. The other parameters are as follows: $k_2 = 4.4$ kg/m$^2$, $m = 0.3822$ kg, $l = 0.1$ m, $P_0 = 10^6$ N, $\omega_* = 3.14$ ms$^{-1}$. For $\lambda_2 = 0.1$ Ns/m, the asymptotics and the results of the numerical experiment agree up to the accuracy of the plotting error. For smaller values of $\lambda_2$, this quantitative agreement holds true in the quasi-front zone only.

## 5. Conclusion

Studying the dynamic behavior of a block-structured medium, we have revealed that the block structure of the medium alters the medium behavior as compared with the continuous medium model obtained by averaging mechanical properties of a discrete medium. In this case, the dynamic behavior of the block medium is well described as the motion of rigid blocks, carried out due to viscous-elastic layers between them.

Using the chain of bars model, we have demonstrated that an impulsive load generates two groups of nonstationary waves in the chain of bars. To be more precise, we have shown that one group of waves consists of low-frequency waves and corresponds to the pendulum wave, while the other group consists of high-frequency waves and corresponds to the fundamental longitudinal ocsillations of the bars.

We have shown that, for approximate description of the low-frequency wave processes, a chain of bars may be replaced by a chain of masses (with masses equal to masses of the bars).

Using both numerical and analytical methods, we have found asymptotic regularities for the attenuation of one-dimensional pendulum waves generated by an impact load in a block medium with viscoelastic interlayers.